\documentclass[aps,prd,reprint,showpacs,amsmath,floatfix]{revtex4-1}
\pdfoutput=1
\usepackage{graphicx}
\usepackage[colorlinks,
    linkcolor=blue,
    citecolor=blue,
    filecolor=blue,
    urlcolor=blue]{hyperref}
\usepackage{enumerate}

\newcommand{\Mao}{Ref.~\citep{2013ApJ...764...35M}}
\newcommand{\MaoF}{Reference~\citep{2013ApJ...764...35M}}
\newcommand{\ExpX}{Exp.~X}
\newcommand{\ExpS}{Exp.~S}

\newcommand{\abs}[1]{{\left\vert{#1}\right\vert}}
\newcommand{\vrms}{v_{\mathrm{rms}}}
\newcommand{\vesc}{v_{\mathrm{esc}}}
\newcommand{\vmin}{v_{\mathrm{min}}}
\newcommand{\sun}{{\odot}}

\begin{document}

\title{Connecting direct dark matter detection experiments \\ to cosmologically motivated halo models}

\author{Yao-Yuan Mao, Louis E.~Strigari, Risa H.~Wechsler}
\affiliation{Kavli Institute for Particle Astrophysics and Cosmology
  and Physics Department, \\Stanford University, Stanford, California 94305, USA\\
  SLAC National Accelerator Laboratory, Menlo Park, California, 94025, USA}

\begin{abstract}
Several direct detection experiments, including recently CDMS-II, have reported signals consistent with 5 to 10 GeV dark matter (DM) that appear to be in tension with null results from XENON and LUX experiments; these indicate a careful review of the theoretical basis, including the galactic DM velocity distribution function (VDF). We establish a VDF parameter space from DM-only cosmological simulations and illustrate that seemingly contradictory experimental results can be made consistent within this parameter space. Future experimental limits should be reported after they are marginalized over a range of VDF parameters.
\end{abstract} 


\maketitle

\section{Introduction}
Dark matter comprises roughly 80\% of the matter in the Universe and seeds the formation of galaxies and large-scale structure. Weakly interacting massive particles (WIMPs) are well-motivated candidates for dark matter (DM), and many theoretical WIMP candidates have been proposed~\citep{1996PhR...267..195J,2005PhR...405..279B,2010pdmo.book.....B,2010ARA&A..48..495F}. Though WIMPs have not been detected, a variety of direct, indirect, and collider experiments are rapidly progressing in searching for them~\citep{2012arXiv1211.7090S}.

Despite rapidly improving sensitivities and analysis methods, direct detection experiments are presenting a conflicting picture. The DAMA~\citep{2010EPJC...67...39B}, CoGENT~\citep{2011PhRvL.106m1301A}, and CRESST~\citep{2012EPJC...72.1971A} Collaborations have reported hints for low-mass DM in the mass range $\sim$~5--10~GeV. Most recently, the CDMS-II Collaboration has reported three events in their silicon detectors that are not explained by known backgrounds.  When interpreted as a WIMP signal this yields a most likely mass of 8.6~GeV~\citep{2013arXiv1304.4279C}. However, these candidate events are inconsistent with the null result reported by the XENON100 Collaboration~\citep{2012PhRvL.109r1301A} and the LUX Collaboration~\citep{Akerib:2013tjd}. Ideas to alleviate the conflict include improved characterization of experimental backgrounds~\citep{2012arXiv1204.3559C,2012PhRvD..86j1301S}, particle physics explanations such as tuning the ratio of the coupling constants of WIMP scattering on neutrons and protons~\citep{2011PhLB..703..124F}, or more detailed examination of the velocity distribution function (VDF)~\citep{2011PhRvD..83b3519L,2013ApJ...764...35M,Bhattacharjee:2012xm,2012JCAP...01..024F}.

The so-called ``standard halo model" (SHM), which assumes a specific value of local DM density and specifies the VDF to be a Maxwell--Boltzmann distribution with a cutoff at the escape velocity, is commonly adopted by direct detection experiments.  As a consequence, uncertainties in the local DM density (see e.g.~\citep{Bovy:2012tw,Garbari:2012ff}) and in the VDF are not a standard component of analysis of experimental data. While the local DM density affects the overall detection rates for all experiments, the VDF affects different experiments differently. For heavy WIMPs, greater than $\sim$ 20~GeV, it is relatively safe to neglect uncertainties in the VDF because the majority of modern experiments are not sensitive to variation of the VDF in this high-mass regime.  However, for lighter WIMPs uncertainties the VDF may significantly affect experimental results.

Cosmological simulations have suggested that DM halos in a Lambda cold dark matter (CDM) universe do not have isothermal profiles~\citep{Navarro:1996gj,Lu:2005tu}, so one does not expect the VDF in DM halos should necessarily follow the isotropic Maxwell-Boltzmann distribution. Recent studies also confirmed this inconsistency by directly comparing the VDFs in simulated halos with the Maxwell-Boltzmann distribution~\citep{2009MNRAS.395..797V, 2010JCAP...02..030K}. VDFs which are consistent with certain anisotropy profiles have been calculated~\citep{Lokas:2000mu,Evans:2005tn}, and parametric VDF models that directly fit to the VDF of simulated halos have also been proposed~\citep{2011PhRvD..83b3519L,2013ApJ...764...35M}.

In addition to the deviation from the Maxwell-Boltzmann distribution due to anisotropy, large substructures or other dark components such as dark discs~\citep{2008MNRAS.389.1041R,Bruch:2008rx} and streams~\cite{2011MNRAS.413.1419V} can result in a nonsmooth VDF that cannot be characterized by the SHM either. Methods to present and compare results from different experiments without assuming a specific VDF model have been developed~\citep{2011PhRvD..83j3514F,2012JCAP...01..024F,2012JCAP...12..015G,Frandsen:2013cna}, though a VDF model is still required to translate results from experiments into constraints or limits on physical parameters of the DM particle~\citep{2009JCAP...11..019S,2010PhRvD..81h7301P,2011PhRvD..83h3505P,2013JCAP...02..041P,2013arXiv1303.6868K,Friedland:2012fa}.

It has not yet become standard in the direct detection community to include uncertainties of the VDF or to use a VDF-independent presentation in published results, possibly because the traditional vanilla WIMP candidate has mass of $\sim 100$~GeV and in this regime experiments are less subject to impact of the VDF. As intriguing signals continue to mount, and new theoretical models of low-mass DM are constructed~\citep{2008PhLB..670...37F,2009PhRvD..79k5016K,2011PhLB..703..124F,2012PhRvD..85g6007E}, it is important to systematically address the issue of the VDF in the context of direct detection experiments. 

Due to our lack of knowledge about the exact form of the VDF, it is not straightforward to include the possible uncertainties in VDF in experimental analyses.  As an initial step, a flexible and parametrized smooth VDF model that is consistent with our understanding of CDM halos is essential at the current stage. In a recent analysis of DM-only cosmological simulations, Mao~\textit{et}~al.~\citep{2013ApJ...764...35M} have empirically determined that the VDF in DM halos may be described by the following functional form with two parameters $(v_0, p)$:
\begin{equation}
  \label{eq:vdf} 
    f(\mathbf{v}; r)\propto
    \begin{cases}
	    \exp\left(-\frac{\abs{\mathbf{v}}}{v_0}\right)
	    \left(\vesc^2-\abs{\mathbf{v}}^2\right)^p,&\abs{\mathbf{v}}\in [0,\vesc]\\
		0,&\mathrm{otherwise},
	  \end{cases}
\end{equation}
where the dependence on $r$ of $v_0$, $p$, and $\vesc$ is omitted in Eq.~(\ref{eq:vdf}) for convenience. Note that the distribution function $f$ in Eq.~(\ref{eq:vdf}) is the distribution of the speed only, not of the full 3-dimensional velocity vector. One cannot construct the distribution of the 3-dimensional velocity from Eq.~(\ref{eq:vdf}) due to the existence of anisotropy. Readers should see Fig.~3 and Sec.~4 of \Mao{} for a detailed discussion on the relation between anisotropy and this distribution function.

This particular functional form is flexible enough to incorporate a wide range of peak velocities and the power-law falloff near $\vesc$. Although it was motivated by DM-only simulations, a recent study shows that this functional form provides an excellent fit to baryonic simulation as well~\citep{Kuhlen:2013tra}. While the baryonic physics impacts the best-fit parameters specifying the VDF, it does not appear to change the general functional form. We use a suite of cosmological simulations and zoom-in simulations to identify a domain of the VDF parameter space that is allowed. We further demonstrate that, within this parameter domain, there exists the intriguing possibility that the tension between these experiments can be resolved by uncertainties in the Milky Way (MW) halo model, and motivates the development of a stronger connection between cosmological simulations and predicted direct detection event rates. We conclude by discussing how this VDF model provides a framework for studying the uncertainties in VDF and suggesting how to mitigate these uncertainties in experimental analyses. 

\section{The Distribution of the VDF Parameters}
\MaoF{} identifies the best-fit VDF parameters $v_0/\vesc$ and $p$ of individual halos from simulations, and indicates an apparent correlation between these two parameters for a fixed $r/r_s$. This degeneracy between $v_0/\vesc$ and $p$ impedes a simple description of the parameter domain of interest. To break this degeneracy, we instead find it useful to parametrize the VDF of Eq.~(\ref{eq:vdf}) by $\vrms/\vesc$ and $p$, where $\vrms$ is the root-mean-square velocity, defined as $\left[4\pi \int_0^{\vesc} dv v^4 f(v) \right]^{1/2}$. For simplicity, hereafter we use $\vrms$ and $v_0$ to refer to their respective normalized values, $\vrms/\vesc$ and $v_0/\vesc$. 

\begin{figure}
\centering \includegraphics{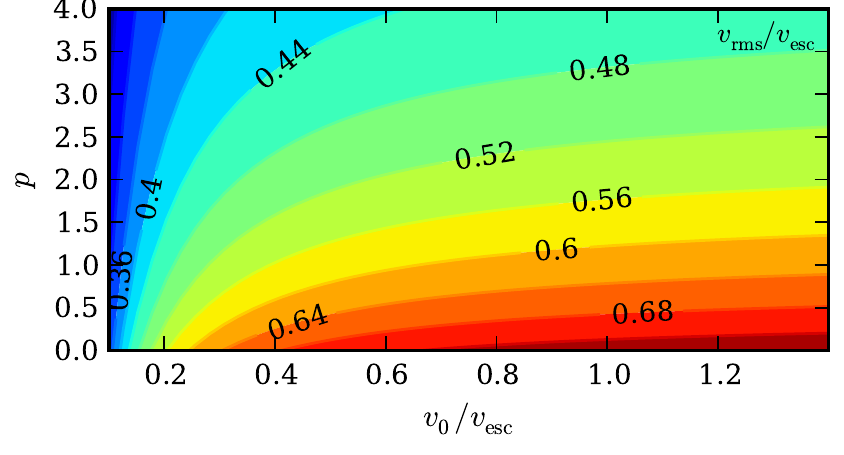} 
\caption{Contours show the value of $\vrms/\vesc$ as a function of $(v_0/\vesc, p)$, from the VDF model of Eq.~\ref{eq:vdf}.} 
  \label{fig:vrms} 
\end{figure}

In Fig.~\ref{fig:vrms} we show the value of $\vrms$ as a function of $(v_0, p)$. There is an one-to-one correspondence between $(\vrms, p)$ and $(v_0, p)$, so the VDF of Eq.~(\ref{eq:vdf}) can be completely specified by $(\vrms, p)$.  Furthermore, lines of constant $\vrms$ follow the relation between $v_0$ and $p$ for a fixed $r/r_s$, where $r_s$ is the scale radius of the density profile; $\vrms$ is largely determined by $r/r_s$, while the halo-to-halo scatter is primarily determined by the parameter $p$.  This is physically explained by noting that $\vrms$ is the ratio of the average energy to the escape energy, which is directly related to the relative position in the gravitational potential.

\begin{figure*}
\centering \includegraphics{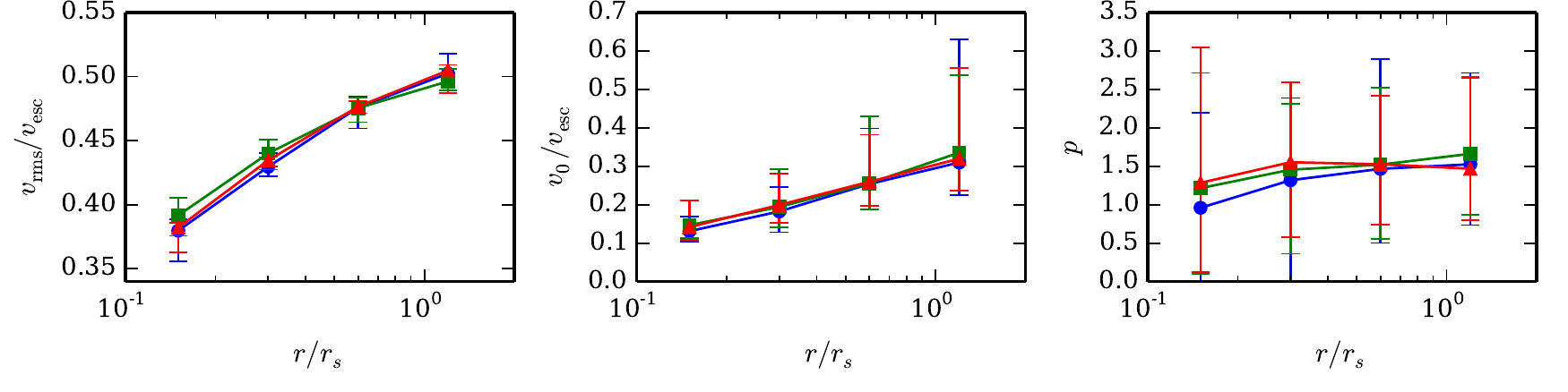} 
\caption{From left to right, plots show $\vrms/\vesc$ (from fitted profiles), fitted $v_0$, and fitted $p$ respectively, as functions of $r/r_s$, for simulated DM halos of three samples.  The red triangles, green squares, and blue circles represent samples of halos of $\sim 10^{13}$, $10^{14}$, and $10^{14.8} M_\sun h^{-1}$, respectively. See text for the simulation detail. Error bars show the 90\% halo-to-halo scatter of each sample.}
  \label{fig:sims} 
\end{figure*}

Figure~\ref{fig:sims} shows the 90\% scatter on the VDF parameters for three different samples of simulated halos. One sample is from the {\sc Rhapsody} simulation~\citep{2013ApJ...763...70W}, in which there are 96 halos with virial mass of $\sim 10^{14.8} M_\sun h^{-1}$. The other two samples are halos with virial mass of $\sim 10^{14} M_\sun h^{-1}$ and of $\sim 10^{13} M_\sun h^{-1}$ respectively, in the the {\sc Bolshoi} simulation~\citep{2011ApJ...740..102K}. We use samples of halos with different masses in order to determine if there are mass trends of the VDF parameters. As shown in \Mao{} and more explicitly in Fig.~\ref{fig:sims}, there is no mass trend indicated over 3 orders of magnitude in mass, implying that it is reasonable to apply the following analysis to MW-mass halos.  

We set the domain of interest on $\vrms$ based on the current observational constraint on $r/r_s$, which is, conservatively, $[0.15,1.2]$~\citep[and references therein]{2012ApJ...761...98K,2013ApJ...764...35M}. This then sets the domain of interest on $\vrms$ to be $[0.35, 0.52]$. Since the parameter $p$ is not affected by $r/r_s$, guided by the 90\% halo-to-halo scatter from Fig.~\ref{fig:sims} we set the domain of interest on $p$ to be $[0, 3]$.  Note that the magnitude of the halo-to-halo scatter is comparable to the directional scatter at a fixed radius within an individual halo, so the above domain will not shrink even if one could remove the halo-to-halo scatter completely, given our lack of knowledge about the Earth's angular position. The simulations used here do not include baryons, so in principle this domain may be larger than what is discussed here.

\section{A Demonstration with Mock Experiments}
We demonstrate the impact of uncertainties in the VDF on direct detection experiments by considering two mock experiments, which we call \ExpX{} and \ExpS{}, and investigate how the different parameters of the VDF in Eq.~(\ref{eq:vdf}) impact the interpretation of the results. In this demonstration, we assume a WIMP model which has a mass $m_\mathrm{dm} = 8.6$~GeV and a WIMP-nucleon cross section at zero momentum transfer $\sigma_0 =1.9\times10^{-41}$~cm$^2$, as inspired by the recent results from the CDMS-II experiment \citep{2013arXiv1304.4279C}.  Note that this mass and cross section are also consistent with the recent CoGENT analysis~\citep{2012PhRvD..85d3515K}.

In \ExpX{}, the target nucleus is xenon, the nuclear recoil energy threshold is 6~keV (i.e.~minimal $\vmin\sim~715$~km/s), and the effective exposure is 6000~kg-days. In \ExpS{}, the target nucleus is silicon, the threshold is 7~keV (i.e.~minimal $\vmin\sim~443$~km/s), and the exposure is 7.1~kg-days, chosen to obtain a mean event count of 3 in the case of the SHM. In both experiments, to highlight the theoretical impact of the VDF we assume a sharp energy cutoff at the threshold energy, and both perfect energy response efficiency and resolution. We fix the local DM density to be $\rho_0 =$ 0.3~GeV$/$cm$^3$, and assume equal WIMP coupling to the neutron and proton. We set the galactic escape velocity to be 544~km$/$s, and take the averaged speed of the Earth in the galactic frame to be 232~km$/$s. Note that we have neglected the uncertainties in $\rho_0$ ($0.3\pm 0.1$~GeV$/$cm$^3$~\citep{Bovy:2012tw}) and $\vesc$~(498--608~km/s at 90 percent confidence~\citep{Smith:2006ym}). In a complete analysis these uncertainties should also be marginalized over.

Given the parameters stated above, we can then calculate the predicted event rate $R$, which is the integral of the differential event rate per unit detector mass over the recoil energy $Q$,
\begin{equation}
  \label{eq:diff_rate}
\left.\frac{dR}{dQ}\right\vert_Q = \frac{\rho_0\sigma_0}{2\mu^2m_\mathrm{dm}}A^2\left\vert F(Q)\right\vert^2\int_{\vmin(Q)}d^3v\,\frac{f(\mathbf{v}+\mathbf{v_e})}{v}.
\end{equation}
Here $\mu$ is the WIMP-nucleon reduced mass, $A$ is the atomic number of the nucleus, $\left\vert F(Q)\right\vert^2$ is the nuclear form factor~\citep{1996APh.....6...87L}, $\vmin=(Q m_N/2\mu^2)^{1/2}$ for an elastic collision, $f$ is the VDF in the galactic rest frame, and $\mathbf{v_e}$ is the velocity of Earth in the galactic rest frame.

The question we address in this demonstration is how the probability of a certain experiment observing $N$ collision events (assuming all the events are real WIMP-nucleus collisions) varies with different models for the VDF. We define $P_X$ to be the probability that \ExpX{} observes \textit{no} events, and $P_S$ the probability that \ExpS{} observes three events. We calculate the probabilities assuming that WIMP-nucleon collision events follow a Poisson process, $P(N;\lambda) = \left(\lambda^N/N!\right) e^{-\lambda}$, where $N$ is the number of events, which equals 0 for $P_X$ and 3 for $P_S$, and $\lambda$ is a dimensionless parameter that equals the predicted rate times the exposure of the experiment. Note that $\lambda$ changes with the WIMP model, the experimental setup, and the VDF.  In the demonstration we always fix the WIMP model and the settings of the two experiments, and only change the VDF to see its effect.

\begin{figure*}
\centering \includegraphics{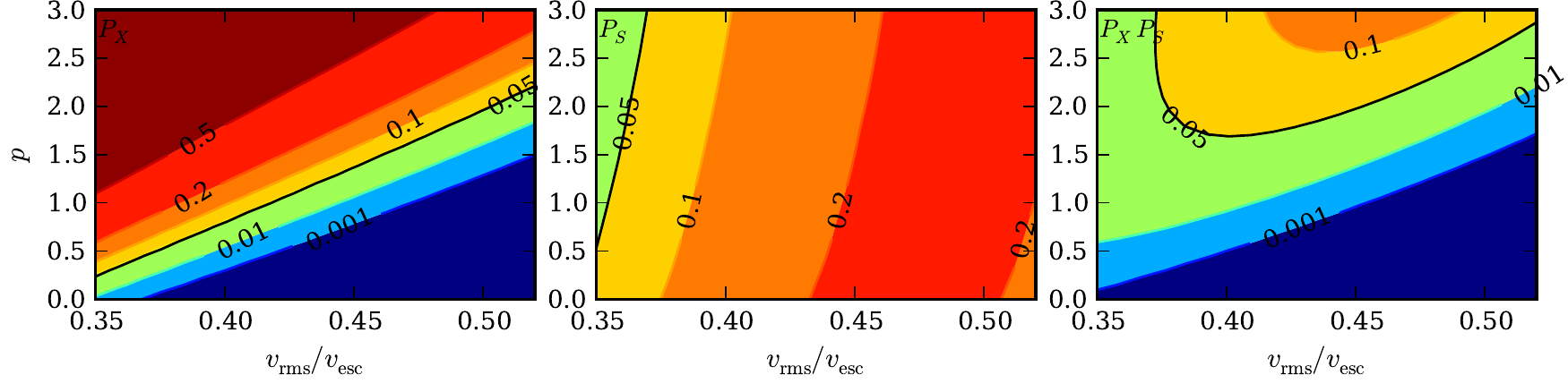}
\caption{Contours show the probabilities $P_X$ (left), $P_S$ (middle), and $P_X\times P_S$ (right), as functions of the VDF parameters $\vrms/\vesc$ and $p$ in the region of interest. The color scale on each panel is the same. $P_X$ is the probability that \ExpX{} observes \textit{no} event, and $P_S$ is the probability that \ExpS{} observes 3 events. Values below 0.05 are excluded with 95\% confidence. High values of $p$ can significantly reduce the tension between the two experiments, when compared to the SHM.}
 \label{fig:likelihood} 
\end{figure*}

Assuming the SHM, we obtain $P_X = 4.65\times10^{-7}$ and $P_S=0.224$. With these assumptions (including the sharp energy cutoff), given the low $P_X$, \ExpX{} rejects the WIMP model at a high confidence level. So if \ExpS{} does indeed observe WIMP events, it implies a strong tension between these two experiments.  Note that when the SHM is assumed, this conflict remains for any escape velocity larger than 515~km$/$s.  However, the results change dramatically if a different VDF model is assumed. Assuming the VDF in Eq.~(\ref{eq:vdf}) with a range of parameters motivated from cosmological simulations, we calculate $P_X$ and $P_S$ and show the results in Fig.~\ref{fig:likelihood}.

The uncertainties in the VDF can have distinct effects on different experiments. Figure~\ref{fig:likelihood} shows that $P_X$ is a strong function of $p$, while $P_S$ only mildly depends on $\vrms$ and is insensitive to $p$. Because different experiments have different responses to changes in the VDF, a given VDF can reconcile two experiments that are inconsistent with one another when using the SHM. 

The leftmost panel of Fig.~\ref{fig:likelihood} shows that \ExpX{}, which is strongly ruled out with the SMH, can only reject less than half of the parameter domain at a 95\% confidence level when the VDF is allowed to vary.  On the other hand, \ExpS{} could still observe three events, given that $P_S > 0.05$ for almost all $\vrms$ and $p$ within the ranges shown on Fig.~\ref{fig:likelihood}.  The rightmost panel shows the joint probability $P_X\times P_S$.  In roughly one-third of the parameter domain, the possibility of \ExpS{} observing three events and \ExpX{} observing none cannot be excluded. To exclude this WIMP model for all possible VDFs considered within this domain at 95\% confidence level, \ExpX{} must lower its energy threshold to at least 5.25~keV, if all other conditions and assumptions unchanged.

The above analysis does not include the effect of background noise, the energy cutoff, the energy response efficiency, and the energy resolution of the mock experiments, and hence caution should be invoked when drawing strong conclusions regarding the relation between XENON100 and CDMS-II experiments. Since the original submission of this manuscript, new results were presented by LUX, and for all values in the VDF parameter space we proposed, the results from LUX and CDMS-II experiments appear to be inconsistent. However, it clearly motivates a full self-consistent statistical analysis with a VDF of the form Eq.~(\ref{eq:vdf}), because if the DM is in fact a light WIMP, a more realistic model for the VDF will be required to translate measurements into physical parameters of the DM particle.

\section{Discussion}
The above demonstration shows that even in a small range of the parameter space of our VDF model, which is consistent with DM-only simulations, the experimental results can already be interpreted very differently. Almost surely there are additional uncertainties which impede a simple choice of the VDF model to be adopted by experimentalists. Nevertheless, in this section we show that the VDF model of \Mao{} provides a framework in understanding these uncertainties.

\MaoF{} presented a detailed discussion of the sources of scatter.  Here we further distinguish these sources according to their contribution to the uncertainties in $\vrms$ or in $p$.  We find here that $\vrms$ is largely determined by $r/r_s$; the uncertainty in this parameter is thus driven by observational uncertainty in $r/r_s$ for the position of the solar system with respect to the density profile of the Milky Way.  Conservative estimates of the concentration parameter of the Milky Way imply the region of $\vrms$ used in Fig.~\ref{fig:likelihood}; with more optimistic assumptions one can constrain $r/r_s \in [0.32, 0.50]$~\citep{2012ApJ...761...98K}.  This will narrow the parameter range shown in Fig.~\ref{fig:likelihood} but would not change our conclusions.  It is likely that future data on the motions of Milky Way halo stars and satellites will be able to further constrain the density profile of our Galaxy's halo to minimize this uncertainty.  

The uncertainty in $p$, on the other hand, at present appears to be irreducible.  The halo-to-halo scatter in $p$ could originate from the different intrinsic properties between halos, but we have not yet found any significant correlations between $p$ and physical properties of the halo (even if found, the quantity may not be well-constrained observationally).  In principle, one could ignore the halo-to-halo scatter if we had a simulation that resembles the Milky Way halo in every way; however, there would still be intrahalo scatter due to variation of VDF in various angular positions at a fixed radius.  In \Mao{}, we found that the intrahalo directional scatter is not smaller than the halo-to-halo scatter. Nevertheless, future measurements of stellar streams and the motions of satellites in the halo of the Milky Way, combined with modeling of large numbers of halos with realistic baryonic physics, could possibly constrain this parameter even in specific regions. Last but not least, baryons could also possibly impact the shape of the VDF as characterized by $p$. For example, one baryonic simulation~\citep{Kuhlen:2013tra} shows a higher value of $p$ ($=2.7$) than in the same halo with DM only, and also has a higher value than the median value we obtained from DM-only simulations.  

\begin{figure}
\centering \includegraphics{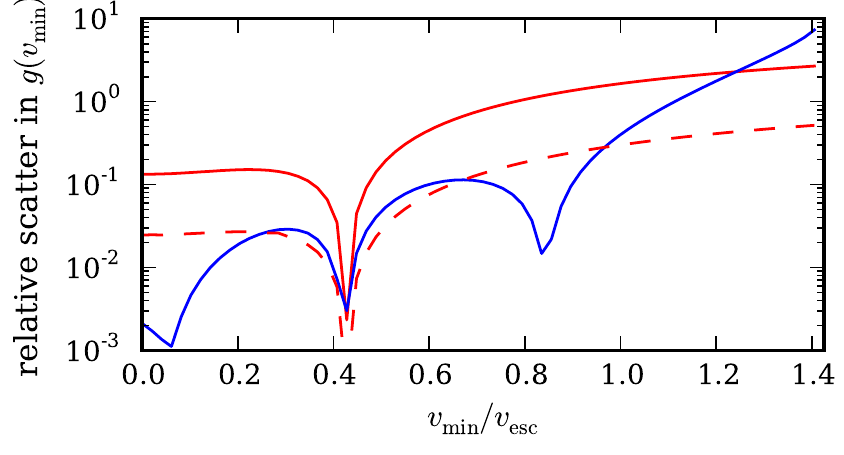}
\caption{Relative scatter in $g(\vmin)$, as defined in text, as a function of $\vmin$. The red solid line shows the effect of $\vrms\in[0.35, 0.52]$ (with $p=1.5$), the red dashed line shows the effect of a reduced parameter space $\vrms\in[0.43, 0.46]$ (with $p=1.5$), and the blue solid line shows the effect of $p\in[0,3]$ (with $\vrms=0.45$). The features (dips) are due to the nonzero speed of the Earth in the galactic frame, and only appear in the scatter of $g(\vmin)$ but not in the energy spectra of the detection experiments.}
  \label{fig:scatter}
\end{figure}

At present it is hence important to include different VDF models or to marginalize over some VDF parameter space, when making statistical statements about signals or exclusions, because different VDF parameters/models that are well within the uncertainties of our current understanding can have very different contribution to the detection rate for different experiments. Figure~\ref{fig:scatter} demonstrates this by showing the relative scatter (defined as the difference between the maximum and the minimum divided by the mean value) in $g(\vmin)$ due to the two parameters defined in Eq.~(\ref{eq:vdf}) for different values of $\vmin$, where $g(\vmin)$ is defined as
\begin{align}
g(\vmin) &\equiv \int_{\vmin} d^3v\,\frac{f(\abs{\mathbf{v}+\mathbf{v_e}})}{v} \\
 &= \frac{2\pi}{v_e} \int_{\max(\vmin-v_e, 0)}^{\vesc} dy \, y L(y) f(y),
 \label{eq:gv_int}
\end{align}
where $L(y) = \min(y + v_e - \vmin, 2y, 2v_e)$ and other variables are defined in the same way as in Eq.~(\ref{eq:diff_rate}). We note that the deduction of Eq.~(\ref{eq:gv_int}) is valid for any generic, smooth or not, VDF model which only depends on the DM speed in the galactic frame.

We note that Eq.~(\ref{eq:vdf}) does not account for all possible astrophysical uncertainties. Nonsmooth components such as dark disks and streams could results in some features in the VDF that cannot be characterized by this model. So far, simulations including hydrodynamics indicate that Eq.~(\ref{eq:vdf}) also fits to the VDF very well in the presence of baryons, but since we have not yet fully understood all the baryonic physics involved, it is possible that these processes can contribute to the VDF in a nontrivial way that has not yet been identified.  Caution should thus be taken when using  Eq.~(\ref{eq:vdf}) to represent the full astrophysical uncertainties.  Nevertheless, for low-mass WIMPs or for heavy-nucleon detectors (i.e.~high $\vmin$), the dominant contribution to the uncertainty of VDF is the power-law falloff near $\vesc$ (and hence also the value of $\vesc$). Equation~(\ref{eq:vdf}) provides a simple yet flexible functional form for this power-law tail, so in the high $\vmin$ regime, the uncertainty in $p$ will change the results most dramatically.

In conclusion, we demonstrate that even when restricting to the cosmologically motivated VDFs discussed herein, a wide range of interpretations remains possible for current experimental results. We should emphasize again that assuming the same halo model does not imply that different experiments are comparable, and our demonstration clearly shows this point. Consequently, to present experimental results, especially to make statistical statements about signals or exclusions, we recommend the following strategies:
\begin{enumerate}[(i)]
\item In the low-mass regime, use a VDF-independent method~\citep{2011PhRvD..83j3514F,2012JCAP...01..024F,Frandsen:2013cna} for several WIMP masses.
\item Show at least two different VDF models to highlight the possible uncertainties. Ideally one should choose two very different ones [e.g.~SHM and the VDF function in Eq.~(\ref{eq:vdf}) with high $p$].
\item Choose a family of VDF model and marginalize over its parameters [$v_0$ and $p$ for Eq.~(\ref{eq:vdf})] and the relevant astrophysical quantities ($\rho_0$ and $\vesc$). In the case of Eq.~(\ref{eq:vdf}), here we provide the priors on its VDF parameters deduced from DM-only cosmological simulations. Future baryonic simulations may change these priors.
\end{enumerate}

\begin{acknowledgments}
This work was supported by the U.S.~Department of Energy 
under Contract No.~DE-AC02-76SF00515 and by a KIPAC Enterprise Grant.
Y.Y.M.~is supported by a Weiland Family Stanford Graduate Fellowship.
We thank Blas Cabrera, Peter Sorensen, and Rafael Lang for useful discussions.  
We also thank Hao-Yi Wu and Oliver Hahn for providing access to the {\sc Rhapsody} simulations, 
Anatoly Klypin and Joel Primack for providing access to the {\sc Bolshoi} simulations, 
and Peter Behroozi for the halo catalogs for both simulations.  
Our work used computational resources at SLAC.
\end{acknowledgments}

\bibliography{vdf2}

\end{document}